\documentclass[useAMS,usenatbib]{mn2e}

\usepackage{verbatim,graphicx,epsfig,dcolumn}
\newcolumntype{.}{D{.}{.}{4}}
\newcolumntype{,}{D{.}{.}{2}}
\newcolumntype{;}{D{.}{.}{1}}
\newcommand{\nodata}{$\cdot\cdot\cdot$}
\newcommand{\lesssim}{{\lower-1.2pt\vbox{\hbox{\rlap{$<$}\lower5pt\vbox{\hbox{$\sim$}}}}}}
\newcommand{\gtrsim}{{\lower-1.2pt\vbox{\hbox{\rlap{$>$}\lower5pt\vbox{\hbox{$\sim$}}}}}}
\def\clap#1{\hbox to0pt{\hss#1\hss}}

\title[NGC 4372: an extended AGB]{On the impossible NGC 4372 V1 and V2: an extended AGB to the \protect{[Fe/H] = --2.2} cluster}
\author[I. McDonald, A. A. Zijlstra, A. Rajoelimanana \& C. I. Johnson]{I.~McDonald$^{1}$\thanks{E-mail: mcdonald@jb.man.ac.uk}, A.~A.~Zijlstra$^{1}$, A.~F.~Rajoelimanana$^{2,3}$, C.~I.~Johnson$^{4,5}$\\
$^{1}$Jodrell Bank Centre for Astrophysics, Alan Turing Building, Manchester, M13 9PL, UK\\
$^{2}$South African Astronomical Observatory, P.O. Box 9, Observatory, 7935, South Africa\\
$^{3}$Astrophysics, Cosmology and Gravity Centre (ACGC), University of Cape Town, Private Bag X3, Rondebosch, 7701, South Africa\\
$^{4}$Department of Physics and Astronomy, UCLA, 430 Portola Plaza, Box 951547, Los Angeles, CA 90095-1547, USA\\
$^{5}$National Science Foundation Astronomy and Astrophysics Postdoctoral Fellow}

\begin{document}

\date{Accepted 9999 December 32. Received 9999 December 32; in original form 9999 December 32}

\pagerange{\pageref{firstpage}--\pageref{lastpage}} \pubyear{9999}

\maketitle

\label{firstpage}

\begin{abstract}
The asymptotic giant branch (AGB) of the globular cluster NGC 4372 appears to extend to unexpectedly high luminosities. We show, on the basis of proper motions and spatial distribution, that the extended AGB is indeed a likely part of the cluster. We also present the first spectra of the very cool (2600 K), very luminous (8000 L$_\odot$), very dusty, oxygen-rich, purported long-period variable stars V1 and V2 that define the AGB tip. In particular, on the basis of their radial velocities, we conclude that V1 and V2 are probably members. We find that V1 and V2 are likely undergoing the superwind phase that terminates their nuclear-burning evolution. We hypothesise that the mass-loss processes that terminate the AGB are inhibited in NGC 4372 due to a lack of atmospheric pulsation and the high gas-to-dust ratio in the ejecta, leading to a delay in the associated enhanced mass loss and dust production. Previously predicted, but never observed, this explains the high mass of the white dwarf in Pease 1 in M15 without the need to invoke a stellar merger. If commonplace, this phenomenon has implications for the mass return from stars, the production of carbon stars and supernovae through the Universe's history, and the AGB contribution to light from unresolved metal-poor populations.
\end{abstract}



\section{Introduction}
\label{IntroSect}

The asymptotic giant branch (AGB) of globular clusters rarely extends more than a magnitude beyond the red giant branch (RGB) tip, either in the near-infrared or in bolometric luminosity \citep{BMvL+09,VFO10}. This is particularly true in low-metallicity clusters (e.g\ \citealt{MvLDB10}), which are typically older \citep{MFAP+09}, with less-massive stars which have lower AGB-tip luminosities \citep{MGB+08}.

\begin{figure}
\centerline{\includegraphics[height=0.50\textwidth,angle=-90]{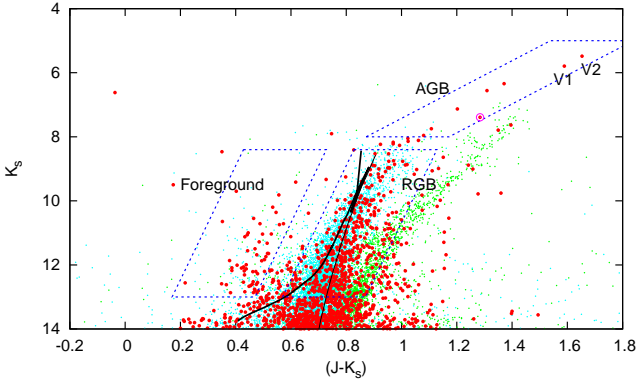}}
\caption{2MASS \protect{($J$--$K_{\rm s})$,$K_{\rm s}$} diagram for stars $<$12.5$^\prime$ from the centre of NGC 4372 (red points). For comparison, $\omega$ Cen (cyan points; [Fe/H] $\approx$ --1.6) and 47 Tuc (green points; [Fe/H] = $-0.7$) are also shown, reddened and scaled appropriately \protect{\citep{Harris96}}. The parallelograms denote regions chosen to represent the cluster's RGB and AGB. The black line shows a [Z/H] = --2.17, 11 Gyr Padova isochrone \protect\citep{MGB+08}, reddened and shifted to match the cluster. The large magenta circle shows the AGB outlier mentioned in Section \ref{AGBSect}.}
\label{CMDFig}
\end{figure}

NGC 4372 is one of the most metal-poor Galactic globular clusters ([Fe/H] = --2.17), and lies at 5.8 kpc distance behind moderate optical obscuration ($E(B-V) = 0.39$ mag) at $l = 301^\circ$, $b = -10^\circ$ \citep{Harris96}. At an estimated age close to that of 47 Tuc and M15 \citep{DAPC+05}, its asymptotic giant branch (AGB) stars initially had a mass of $\approx$0.89 M$_\odot$ \citep{MBvL+11}, placing the AGB and RGB tips at similar luminosity. Nevertheless, there are several stars toward the cluster which appear to form an AGB extending $\approx$3 $K_{\rm s}$ magnitudes above the RGB tip (Figure \ref{CMDFig}). V1 and V2 are variable stars in the cluster's vicinity \citep{CMD+01}, and appear to form the AGB tip, but nothing appears published about these stars except photometry from all-sky surveys. We herein seek to confirm both their membership and the presence of an extended AGB.

\section{V1 and V2}
\label{VarSect}

\subsection{Observations}
\label{ObsSect}

\begin{figure}
\centerline{\includegraphics[height=0.50\textwidth,angle=-90]{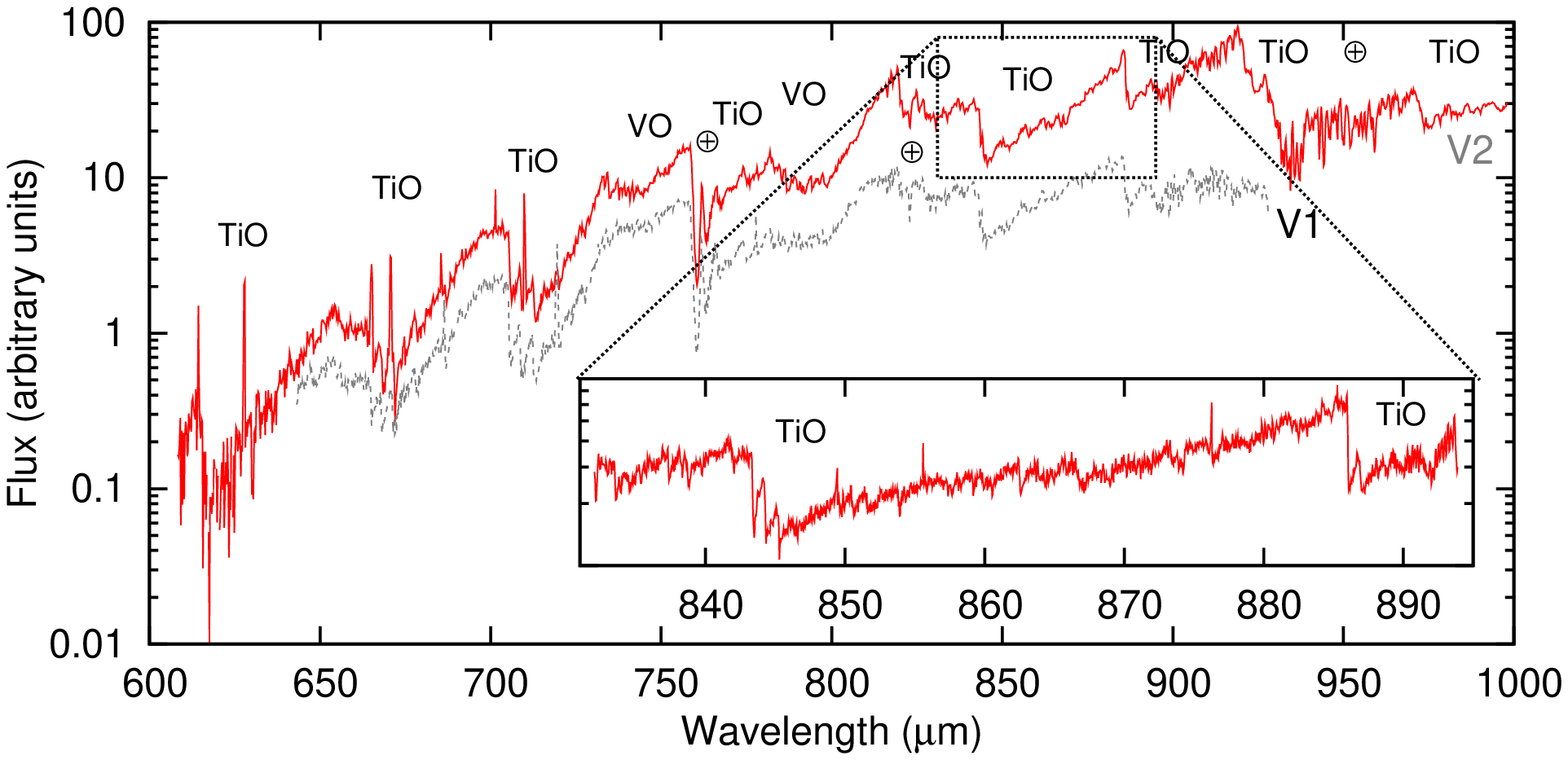}}
\centerline{\includegraphics[height=0.50\textwidth,angle=-90]{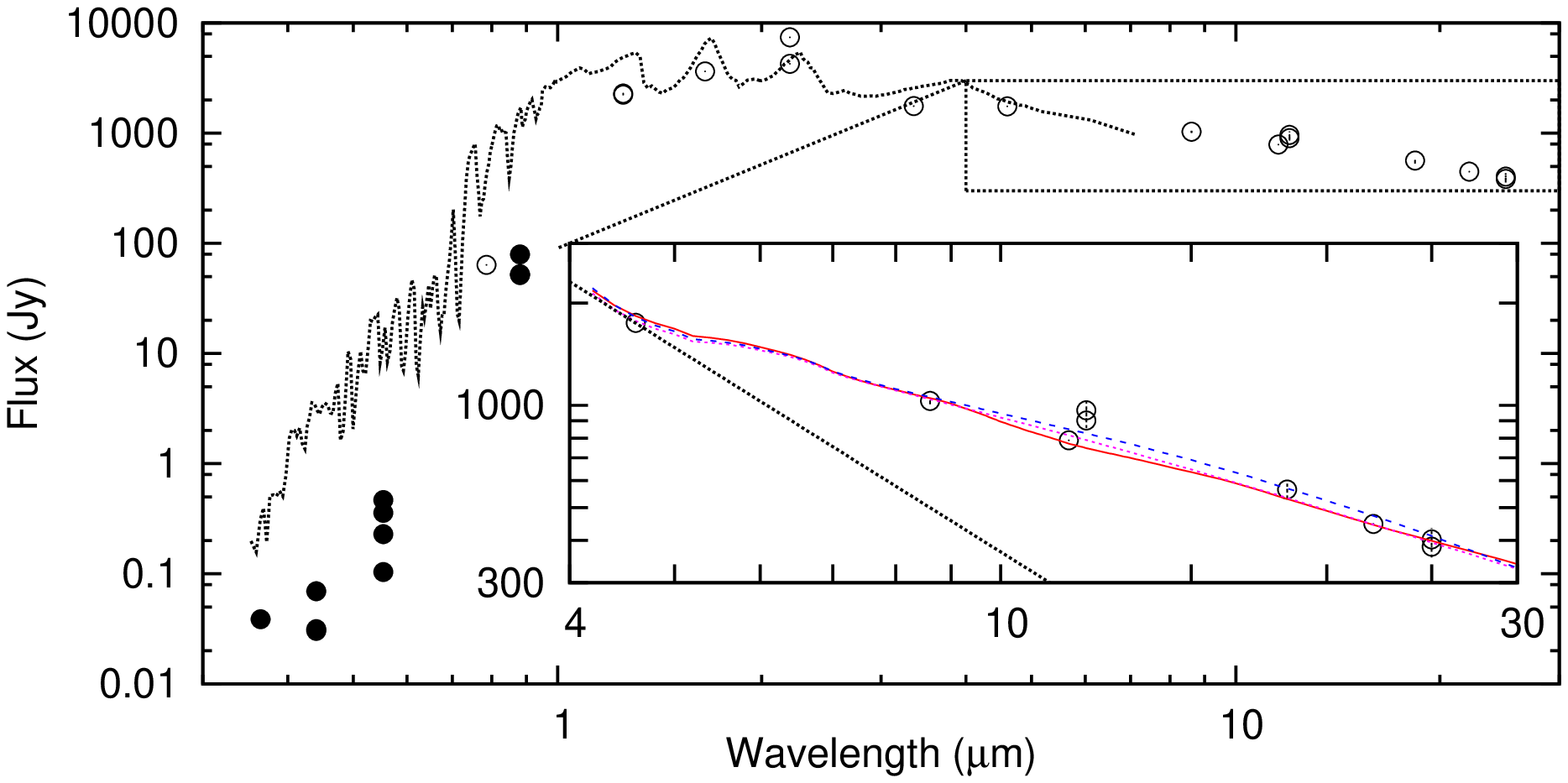}}
\caption{Top panel: spectra of V1 and V2 from the SAAO. Bottom panel: SED of V2. The hollow points show literature data from DENIS \citep{EdBC+94}, 2MASS \citep{SCS+06}, \emph{WISE} \citep{CWC+12}, \emph{AKARI} \citep{KAC+10}, and \emph{IRAS} \citep{MCC+90}. The solid points show magnitudes estimated from archival ESO 2.2-m/WFI observations. The black dotted line is a 2600 K, $\log(g) = 0$, [Fe/H] = --2 \protect{\sc BT-Settl} model atmosphere, reddened using $E(B-V) = 0.39$ mag. The inset shows {\sc dusty} model fits for metallic iron (magenta, dotted line), amorphous carbon (blue, dashed line), large silicate grains (red, solid line).}
\label{SEDFig}
\end{figure}

To determine the nature of V1 and V2, we took optical spectra using the grating spectrograph on the 1.9-m Radcliffe Telescope at SAAO. The spectrum of V1 was taken on 2012 Apr 17 and covers 643--928 nm at a resolution of $R = \Delta\lambda/\lambda \approx 4800$. The low-resolution spectrum of V2 was taken on 2012 Apr 16, covering 608--998 nm at $R \approx 3600$; the high-resolution spectrum on 2012 Apr 13, covering 832--894 nm at $R \approx 24\,000$. We show these spectra along with literature photometry in the top panel of Figure \ref{SEDFig}.

Both stars are clearly oxygen rich and very cool, showing TiO and VO bands: we estimate the spectral type of V1 and V2 to be M7--M8 and M8--M9, based on the Fluks spectral standards \citep{FPT+94}. These bands obscure the atomic lines, making accurate radial velocity determination difficult. The heliocentric radial velocity for V2 is found to be $v_{\rm r}$ = 100 $\pm$ 10 km s$^{-1}$ based on comparison to atmospheric lines, the CuNe arc lamp spectrum and subsequent observation of the M0 star HIP 56619. The low-resolution of the V1 spectrum precluded an accurate radial velocity measurement: we estimate it lies between $v_{\rm r} = 50$ and 100 km s$^{-1}$. These velocities are significantly different from that of the contaminating Galactic foreground population, which we measure to be 3 $\pm$ 21 km s$^{-1}$ (standard deviation) on the basis of an initial reduction of archival VLT/FLAMES data of the cluster's RGB stars.

As no variability information (and precious little reliable optical data) is available in the literature, we obtained archival imagery taken with the ESO 2.2-m/WFI at La Silla. This contains $UBVI_{\rm C}$-band data with four epochs between  2001.157--2002.466. The variability between epochs was obtained by comparison to nearby stars. We find the variability of V2 to be $\Delta B,V,I_{\rm C} > 0.85$, 0.81 and 0.46 mag, respectively. For V1, we find that $\Delta B,V,I_{\rm C} > 0.67$, 0.49 and 0.23 mag, respectively. There is insufficient data to constrain any period, which may be longer than the temporal coverage. Most likely, these values have been underestimated by a factor 1.5, but our lower limits are (respectively) thrice and twice that found in the slightly-lower metallicity star K825 in M15 \citep{MvLDB10}.

We show our estimated optical magnitudes along with literature infrared data in the lower panel of Figure \ref{SEDFig}. It is clear that the infrared photometry does not follow a Rayleigh--Jeans tail, but exhibits an infrared excess. We have attempted to model this excess using the {\sc dusty} radiative transfer code \citep{NIE99} for metallic iron, amorphous carbon and large silicate grains by assuming a radiatively-driven wind (Figure \ref{SEDFig}, bottom panel). It is clear that any of the three dust species can be made to model a featureless modified blackbody, though we cannot rule out (without 10-$\mu$m spectroscopy) the presence of a small silicate feature in either emission or absorption. Assuming a luminosity of 8000 L$_\odot$, all three models require mass-loss rates of $\dot{M} \gtrsim 10^{-5} \sqrt{(\psi/30\,000)}$ M$_\odot$ yr$^{-1}$, or expansion velocities of $v_{\infty} \lesssim 0.5 \sqrt{(30\,000/\psi)}$ km s$^{-1}$, where $\psi$ is the gas-to-dust ratio, which we take to be 30\,000 at the metallicity of the cluster. While these values may not be accurate in an absolute sense (see discussion in \citealt{MBvLZ11}), the mass-loss rate appears to be $\sim$10$\times$ larger than the highest mass-loss rates in other clusters, which have been computed using similar methods \citep{MBvLZ11,MvLS+11}.

\subsection{Cluster members?}
\label{MemSect}

\begin{figure}
\centerline{\includegraphics[height=0.50\textwidth,angle=-90]{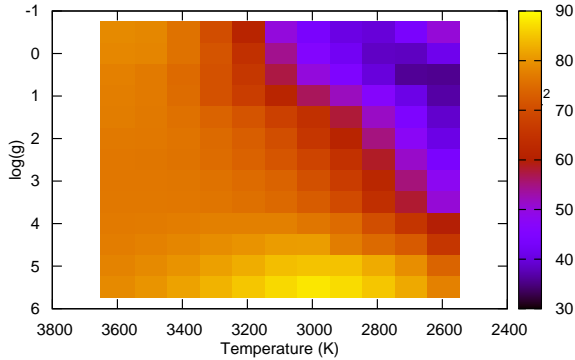}}
\caption{Reduced $\chi^2$ for fits between the spectrum of V2 and \protect{\sc BT-Settl} model spectra at [Fe/H] = --2. The colour scale shows the $\chi^2$ goodness-of-fit with darker colours being better matches.}
\label{Chi2Fig}
\end{figure}

The late spectral types of V1 and V2 force them to be foreground dwarfs, optically-enshrouded young stellar objects (YSOs), or giant stars. Their radial velocity, variability, and lack of strong 7665 and 7699 \AA\ K {\sc i} lines preclude either dwarfs or YSOs. While the dark nebula Sandqvist 149 (the `Dark Doodad'; \citealt{Sandqvist77}) lies near the cluster, it is not actively star-forming nor gives extra extinction to these stars compared to the rest of the cluster. Its distance is unknown, but is likely to be $<$500 pc, based on a Galactic molecular gas scale height of 81 pc \citep{Cox05}. The stellar pulsations suggest a luminosity of $\gtrsim$680 L$_\odot$ \citep{MZB12}, placing their distance from the Sun at $\gtrsim$1.5 kpc, $\gtrsim$260 pc below the Galactic Plane. This supports the idea that they are distant giants lying out of the Galactic Plane.

V1 and V2 are extra-ordinarily cool for stars which we suppose are at [Fe/H] = --2.17 \citep{Harris96}. Strong molecular bands preclude us from using the normal indicators to determine their metallicity: the Ca\,{\sc ii}, Fe\,{\sc i} and Ti\,{\sc i} lines normally visible in cool star spectra are practically indistinguishable from the molecular absorption, though this may be expected from a metal-poor star. This makes it difficult to use conventional techniques to estimate the stellar parameters, thus we use three different {\it ad hoc} methods.

Firstly, we simply note that R Leo provides a close match to both the detailed spectra and broad SED of V2. R Leo has a luminosity of 8200 L$_\odot$ and is at a distance of 110 $\pm$ 9 pc \citep{FWP+05,RM07}. Scaling R Leo to match the near-IR flux of V2 implies a distance of 5 $\pm \approx 1$ kpc (cf.\ NGC 4372 at 5.8 kpc).

Secondly, we compare the stellar SEDs to {\sc BT-Settl} and {\sc marcs} model spectra \citep{GBEN75,AGL+03,GEE+08} using the code of \citet{MvLD+09}. Assuming both stars are at 5.8 kpc, we find that $L_{\rm V1,\ V2} \sim 7000$, 8000 L$_\odot$. These values are subject to a substantial error as the derived stellar temperatures ($T_{\rm V1,\ V2} \sim 2300,\ 2000$ K) are outside the range of validity of the models (2600 and 2500 K, respectively). Simple trapezoidal integration of the SED, however, suggests the luminosities are no more than 15 per cent in error.

Finally, we perform a $\chi^2$ fit between the normalised observed and model spectra in the range 8470--8930 \AA. A strong degeneracy exists between temperature and surface gravity, and we are again limited by the range of the models, thus we cannot fully constrain the parameters. Furthermore, neither model set correctly reproduces the observed VO bands. For V2, the best-fitting {\sc BT-Settl} model at [Fe/H] = --2 (Figure \ref{Chi2Fig}) is at the lowest temperature ($T = 2600$ K) with $\log(g) = 0.5$. The (incomplete) grid of {\sc marcs} models shows a poorer fit at 2500 K than 2600 K. Based on these values, we therefore estimate the parameters of V2 as $L = 8000$ L$_\odot$, $T = 2550$ K, $R = 460$ R$_\odot$, $\log(g) = -1.1$ and $v_{\rm esc} = 23$ km s$^{-1}$. The best-fit match for V2 gives $L = 7000$ L$_\odot$, $T = 2700$ K, $R = 380$ R$_\odot$, $\log(g) = -0.9$ and $v_{\rm esc} = 26$ km s$^{-1}$. All these values are subject to substantial but presently-unquantifiable errors, but are consistent with parameters expected for AGB stars of that luminosity.

However, if V1 and V2 are members, we are left with a slight puzzle: how to reconcile the radial velocity of V2 (100 $\pm$ 10 km s$^{-1}$) with that of the cluster (72.3 $\pm$ 1.2 km s$^{-1}$; \citealt{Harris96}; dispersion 4.6 $\pm$ 0.5 km s$^{-1}$; \citealt{GPCM95}). One possible solution is radial pulsation: for example, the change in observed radial velocity caused by pulsation excceds $\pm$10 km s$^{-1}$ for the strongest variables in 47 Tuc \citep{LWH+05} and is $\pm$13.5 km s$^{-1}$ for the CO bands of R Leo \citep{Hinkle78}. This could bring the discrepency within the error budget. Binarity might also explain the difference, though globular cluster binary fractions are typically very low \citep{IBFR05}.


\section{The extended AGB}
\label{AGBSect}

\begin{figure}
\centerline{\includegraphics[height=0.50\textwidth,angle=-90]{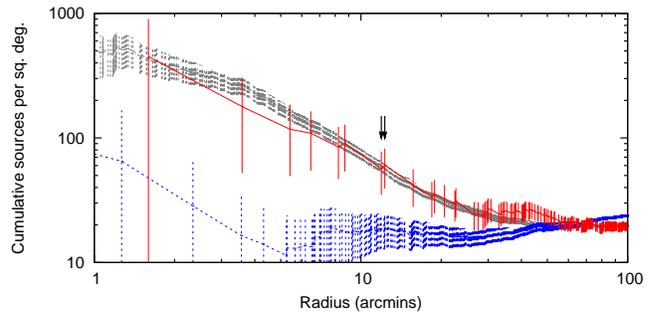}}
\caption{Source density within a given radius of NGC 4372, showing AGB stars (red line), RGB stars (grey line, scaled down by a factor of 30) and foreground control field (blue line, scaled down by a factor of 10), using the boundaries defined in Figure \ref{CMDFig}. Vertical lines show Poissonian errors on the number of stars. V1 and V2 are indicated by arrows, though they lie at opposite sides of the cluster.}
\label{RadialFig}
\end{figure}

Using the definitions in Figure \ref{CMDFig}, the extended AGB within 12.5$^\prime$ is comprised of seven stars (listed in Table \ref{StarsTable}) plus one outlier which we do not consider further. The optical and infrared colours are consistent with a sequence of progressively redder giant stars. This sequence starts, as expected, from the RGB tip. The Padova isochrone in Figure \ref{CMDFig} shows the RGB tip to lie at $K_{\rm s} \approx 8.4$ mag. The isochrone also follows stars through the thermal-pulsating AGB (TP-AGB), but does not correctly model the proposed upper AGB. We could not fit any other isochrone from \citet{MGB+08} to this sequence. Stars above $\approx$0.9 M$_\odot$ should become carbon-rich at this metallicity (see also \citealt{KL07,MG07}). Since V1 and V2 are oxygen-rich, we cannot expect younger (higher-mass) metal-poor isochrones to correctly reproduce a sequence of oxygen-rich stars.

The proposed upper AGB, as defined in Figure \ref{CMDFig}, shares a radial density distribution with the RGB, but not the foreground population (Figure \ref{RadialFig}). In combination with the radial velocity limits placed on V1 and V2, this argues that these stars are indeed part of the cluster.

Proper motions of cluster members should typically be $\lesssim$1 mas yr$^{-1}$ from the cluster mean (cf.\ \citealt{vLlPR+00}). While the proper motions listed in Table \ref{StarsTable} appear greater, the formal errors are likely under-estimated in this crowded field: the RGB population in UCAC4 and SPM4, for example, has a mean and standard deviation of $\Delta\alpha = -2 \pm 30, \Delta\delta = 3 \pm 31$ and $\Delta\alpha = -3 \pm 21, \Delta\delta = 2 \pm 20$ mas yr$^{-1}$, respectively. Only star \#4 shows a non-zero proper motion that is consistently among the catalogues, and even that is well within the standard deviations listed above.

\begin{center}
\begin{table}
\caption{Data for the candidate upper AGB stars. Photometry comes from APASS \citep{HLT+12}; DENIS \citep{EdBC+94}; 2MASS \citep{SCS+06}; and \emph{WISE} \citep{CWC+12}. Proper motions are shown from PPMXL \citep{RDS10}; UCAC4 (Zacharias et al., in prep.) and SPM4 \citep{GvAZ+11}.}
\label{StarsTable}
\begin{tabular}{@{}l@{\ }c@{\ }c@{\ }c@{\ }c@{}r@{\,$\pm$\,}l@{,}r@{\,$\pm$\,}l@{}}
    \hline \hline
2MASS		& \multicolumn{4}{c}{Photometry (mag)}& \multicolumn{4}{c}{\clap{Proper motion (mas/yr)}}\\
RA \& Dec	& $B$	& $V$	& $R$	& $I$		& \multicolumn{4}{c}{PPMXL}\\
(J2000.0)	& $J$	& $H$	& $K$	& \ 		& \multicolumn{4}{c}{UCAC4}\\
Radius ($^\prime$)& [3.35]& [4.60]& [11.6]& [22.1]	& \multicolumn{4}{c}{SPM4}\\
    \hline
12 28 05.686	& 16.040 & 15.470 & 13.465 & 11.455	&    0.7 & 11.8 &  --3.1 & 11.8\\
--72 45 19.94	&  7.136 &  6.125 &  5.483 & \ 		&    2.5 &  4.0 &  --4.5 &  4.0\\
11.92 (V2)	&  5.610 &  4.980 &  4.010 & 3.180	&  --5.7 &  2.8 &    2.1 &  3.1\\
    \hline
12 23 00.611	& 17.085 & 15.248 & 14.490 & 10.227	&    3.9 & 12.3 & --19.9 & 12.3\\
--72 40 03.04	&  7.383 &  6.307 &  5.794 & \ 		& --18.3 &  4.1 &  --9.1 &  4.0\\
12.29 (V1)	&  5.545 &  5.094 &  4.520 & 3.945	&   14.6 &  1.6 &   12.4 &  1.7\\
    \hline
12 25 22.538	& 15.030 & 13.052 & 12.195 & 9.871	& --17.8 & 10.1 &    5.2 & 10.1\\
--72 33 17.48	&  7.771 &  6.744 &  6.340 & \ 		& --10.9 &  2.9 &  --0.8 &  1.8\\
6.48 (\#3)	&  6.207 &  6.349 &  6.140 & 5.984	&\nodata&\nodata&\nodata&\nodata\\
    \hline
12 24 57.740	& 14.144 & 12.096 & 11.273 & 9.831	&    5.5 &  9.1 &   24.6 &  9.1\\
--72 39 04.98	&  7.866 &  6.917 &  6.557 & \ 		&    5.3 &  1.6 &   12.1 &  1.6\\
3.58 (\#4)	&  6.457 &  6.572 &  6.418 & 6.258	&    0.4 &  1.3 &   16.0 &  1.3\\
    \hline
12 25 27.364	& 13.740 & 11.877 & 11.132 & 9.913	&  --6.6 & 11.8 & --26.5 & 11.8\\
--72 38 41.41	&  8.333 &  7.416 &  7.131 & \ 		&  --9.5 &  1.6 &  --5.1 & 10.6\\
1.60 (\#5)	&  6.987 &  7.069 &  6.979 & 6.826	&  --5.6 &  1.3 &  --4.6 &  1.4\\
    \hline
12 27 19.786	& 14.157 & 12.374 & 11.662 & 10.399	&  --4.7 & 10.1 &    3.0 & 10.1\\
--72 35 17.53	&  8.858 &  8.017 &  7.750 & \ 		&    3.0 &  1.4 &  --3.8 &  1.5\\
8.23 (\#6)	&  7.614 &  7.703 &  7.587 & 7.445	&  --1.9 &  1.0 &  --0.9 &  1.0\\
    \hline
12 24 49.536	& 14.081 & 12.275 & 11.580 & 10.355	&  --2.6 & 11.6 & --20.5 & 11.6\\
--72 43 00.74	&  9.036 &  8.203 &  7.956 & \ 		&  --5.3 &  1.5 &  --7.9 &  1.9\\
5.41  (\#7)	&  7.778 &  7.771 &  7.644 & 7.454	&  --5.0 &  1.1 &    4.9 &  1.2\\
    \hline
\end{tabular}
\end{table}
\end{center}


\section{Discussion}
\label{DiscSect}


Of the 13 Galactic globular clusters with [Fe/H] $<$ --2 \citep{Harris96}, only NGC 4372 has a visually obvious extended AGB. Many are sparsely populated enough that an extended AGB would be a stochastic phenomenon, with only one or two stars. Other extended AGBs may therefore have been missed. However, we note that the most-metal-poor globular cluster (M15; [Fe/H] = --2.37; \citealt{Harris96}) unusually contains a planetary nebula, Pease 1, with a white dwarf of 0.58--0.62 M$_\odot$ \citep{BFB+95,ABL00,BBC+01}. This is rather higher than the 0.53 M$_\odot$ found in other clusters \citep{KSDR+09}, but consistent with the core mass estimated for NGC 4372 V2 \citep{Blocker93}. Some stars may therefore reach unexpectedly-high luminosities in M15 too. However, NGC 4372 may simply be unique.

We consider four ways an extended AGB could occur: (1) a young, presumably metal-rich sub-population with higher mass (cf.\ $\omega$ Cen; \citealt{LJS+99}); (2) a sub-population arising from evolved stellar mergers; (3) excess luminosity above the canonical core-mass--luminosity relation (cf.\ hot bottom burning; HBB; \citealt{Herwig05}); (4) reduced mass loss during the star's evolution.

There are four stars on the AGB between $8.0 < K_{\rm s} < 7.0$ mag and 26 stars on the RGB between $9.4 < K_{\rm s} < 8.4$ mag, giving a drop in source density of a factor of 6.5 $\pm$ 3.5 across the RGB tip. Other magnitude ranges give similar results. While Poisson noise is significant, this is commensurate with the factor of $\sim$4 drop expected if all RGB stars survive onto the upper AGB (cf.\ \citealt{SB04}) and suggests that population giving rise to the extended AGB comprises $\gtrsim$40 per cent of the cluster. While the main sequence is broadened, this is attributable to differential reddening over the cluster \citep{ALAW91}, rather than an age or metallicity spread within it. To yield 0.62 M$_\odot$ white dwarfs, such a population would also have to be exceptionally massive ($\sim$2 m$_\odot$) and very young ($\sim$1 Gyr) to reach these luminosities \citep{Blocker93,SSWMB09}. This negates case (1).

The drop in sources over the RGB tip also contradicts case (2), which can be rejected due to the cluster's low density. NGC 4372 has one of the lowest central concentrations of stars of any globular cluster and is not core collapsed, resulting in fewer stellar encounters \citep{Harris96}. While NGC 4372 does show an extended blue horizontal branch (HB) or blue straggler population (signs of stellar mergers), these stars are very few in number compared to the total cluster \citep{KK93}. In order for $\gtrsim$40 per cent of the cluster to reach this phase, at least four of every seven stars must merge. We therefore rule out case (2).

The temperatures required for HBB should not occur at all in stars below $\lesssim$1.5 M$_\odot$ \citep{VM10}. We therefore do not consider case (3) a likely option.

We are therefore left with case (4). The conventional mechanisms by which oxygen-rich AGB stars lose mass, already strained at solar metallicity \citep{Woitke06b} become even more difficult at low metallicity: the pulsations levitating material from the stellar surface are typically weaker, the dust that condenses in the outflow forms a smaller fraction of the wind, meaning that stellar radiation pressure has more difficulty accelerating both it and (by collisional coupling) the surrounding gas from the star. It has been suggested \citep{BW91,Zijlstra04} that, at low metallicity, the inability to effectively drive a dust-driven wind is sufficient to decrease mass-loss rates substantially. Stars are then given more time to ascend the asymptotic giant branch (AGB) before an enhanced superwind ejects the remainder of the envelope into the interstellar medium (ISM).

The blue nature of the HB \citep{KK93} suggests that pre-AGB mass loss occurs normally. The remaining envelope ($\approx$0.1 M$_\odot$; \citealt{MJZ11}) should mostly be lost via pulsation-enhanced, dust-driven winds at luminosities above $\sim$700 L$_\odot$ \citep{MvLS+11,MZB12}. However, we have seen that dust is not produced in great quantity by any stars except V1 and V2 (Table \ref{StarsTable}), and V1 and V2 appear to be the only long-period variable pulsators (N.~Matsunaga, private communication). We therefore theorise that the lack of pulsation and dust formation delays the usual dust-driven wind until the stars reach much cooler temperatures.

If this is typical of very-low-metallicity populations, this would have significant consequences. The increased AGB lifetime would lead to an increased white dwarf mass and decreased return of metal-rich material to the ISM. That similar extended AGBs are not seen in other clusters of similar metallicity may point to detailed elemental abundances being more important in iron content.


\section{Conclusions}
\label{ConcSect}

We provide evidence that mass loss in NGC 4372 is delayed due its low metallicity, explaining the extension of its AGB to $\approx$8000 L$_\odot$. We present spectra of the two unusually cool and dusty stars, V1 and V2, which form the AGB tip. We also give evidence to associate the extended AGB, and V1 and V2 in particular, with the cluster. While NGC 4372 thus far remains unique, we reconcile its extended AGB with the existence of the planetary nebula within M15. We finish by discussing the considerable impact an extended AGB has on stellar evolution, our view of the early Universe and its chemical evolution to date.


\section*{Acknowledgements}

We are very grateful to the staff at SAAO for co-ordinating the observations that made this paper possible. Based on observations made with ESO telescopes at La Silla Paranal observatory under programmes 068.D-0265, 069.D-0582, 088.D-0026, 164.O-0561 and 188.B-3002. This material is based upon work supported by the National Science Foundation under award No.\ AST-1003201 to CIJ.


\label{lastpage}

\end{document}